\documentclass[aps,prd,a4paper,onecolumn,amsmath,showpacs,superscriptaddress,nofootinbib,preprintnumbers]{revtex4-1}
\pdfoutput=1
\usepackage{amssymb,amsmath,latexsym,mathrsfs}
\usepackage{graphicx,subfigure}
\usepackage{epsfig}
\usepackage{varioref,xr-hyper}
\usepackage{color}
\usepackage{multirow}
\usepackage{array}
\usepackage{hyperref}
\usepackage{wasysym}
\usepackage{color}
\usepackage{float}
\usepackage{xcolor}
\usepackage[utf8]{inputenc}
\usepackage[T1]{fontenc}
\begin{document}

\newcommand{\EDV}[1]{\textcolor{blue}{{ [EDV: \bf#1]}}}

\title{Tale of stable interacting dark energy, observational signatures, and the $H_0$ tension}

\author{Weiqiang Yang}
\email{d11102004@163.com}
\affiliation{Department of Physics, Liaoning Normal University, Dalian, 116029, P. R. China}

\author{Supriya Pan}
\email{span@research.jdvu.ac.in}
\affiliation{Department of Mathematics, Raiganj Surendranath Mahavidyalaya, Sudarshanpur, Raiganj, Uttar Dinajpur, West Bengal 733134, India}

\author{Eleonora Di Valentino}
\email{eleonora.divalentino@manchester.ac.uk}
\affiliation{Jodrell Bank Center for Astrophysics, School of Physics and Astronomy, University of Manchester, Oxford Road, Manchester, M13 9PL, UK}

\author{Rafael C. Nunes}
\email{rafadcnunes@gmail.com}
\affiliation{Departamento de F\'isica, Universidade Federal de Juiz de Fora, 36036-330, Juiz de Fora, MG, Brazil}

\author{Sunny Vagnozzi}
\email{sunny.vagnozzi@fysik.su.se}
\affiliation{The Oskar Klein Centre for Cosmoparticle Physics, Stockholm University, Roslagstullbacken 21A SE-106 91 Stockholm, Sweden}
\affiliation{The Nordic Institute for Theoretical Physics (NORDITA), Roslagstullbacken 23, SE-106 91 Stockholm, Sweden}

\author{David F. Mota}
\email{d.f.mota@astro.uio.no}
\affiliation{Institute of Theoretical Astrophysics, University of Oslo, P.O. Box 1029 Blindern, N-0315 Oslo, Norway}

\begin{abstract}
We investigate the observational consequences of a novel class of stable interacting dark energy (IDE) models, featuring interactions between dark matter (DM) and dark energy (DE). In the first part of our work, we start by considering two IDE models which are known to present early-time linear perturbation instabilities. Applying a transformation depending on the dark energy equation of state (EoS) to the DM-DE coupling, we then obtain two novel stable IDE models. Subsequently, we derive robust and accurate constraints on the parameters of these models, assuming a constant EoS $w_x$ for the DE fluid, in light of some of the most recent publicly available cosmological data. These include Cosmic Microwave Background (CMB) temperature and polarization anisotropy measurements from the \textit{Planck} satellite, a selection of Baryon Acoustic Oscillation measurements, Supernovae Type-Ia luminosity distance measurements from the JLA sample, and measurements of the Hubble parameter up to redshift $2$ from cosmic chronometers. Our analysis displays a mild preference for the DE fluid residing in the phantom region ($w_x<-1$), with significance up to 95\% confidence level, while we obtain new upper limits on the coupling parameter between the dark components. The preference for a phantom DE suggests a coupling function $Q<0$, thus a scenario where energy flows from the DE to the DM. We also examine the possibility of addressing the $H_0$ and $\sigma_8$ tensions, finding that only the former can be partially alleviated. Finally, we perform a Bayesian model comparison analysis to quantify the possible preference for the two IDE models against the standard concordance $\Lambda$CDM model, finding that the latter is always preferred with the strength of the evidence ranging from positive to very strong.
\end{abstract}


\maketitle
 
\section{Introduction}

In the concordance $\Lambda$CDM model, dark matter (DM) and dark energy (DE) are pictured as non-interacting components. Nonetheless, cosmological models where DM and DE interact with each other through some dark couplings exist, historically motivated by the possibility of addressing the cosmic coincidence problem, although it is now understood that the background energy exchange (and corresponding couplings) necessary to address such problem is too large and observationally excluded (see e.g.~\cite{DE_DM_1,DE_DM_2} for reviews). In addition, such interactions were initially also considered in order to alleviate the cosmological constant problem~\cite{Wetterich:1994bg}. Such classes of models are usually referred to as coupled or interacting dark energy (IDE) models.

In recent years, interacting dark energy models, as well as similar related models, have been studied within a plethora of contexts: these range from model-building, to simulations, and finally to observational tests~\cite{Amendola:1999er, Billyard:2000bh,Barrow:2006hia,delCampo:2008jx, Quartin:2008px,Valiviita:2008iv,Majerotto:2009np, Honorez:2010rr,Clemson:2011an,skordis,akrami, barrow,winther,Pan:2013rha,Yang:2014gza, Pan:2014afa, yang:2014vza,Faraoni:2014vra, Pan:2012ki,Duniya:2015nva,Tamanini:2015iia,Murgia:2016ccp,Marcondes:2016reb,Yang:2016evp,Mukherjee:2016lor,Pan:2016ngu,Mukherjee:2016shl,Erdem:2016hqw,Sharov:2017iue,Pan:2017ent,Yang:2018pej,Nunes:2016dlj, Q1, Q2, Q3, Q4,Q5,Q6,Otalora:2013tba,Baldi:2011qi,
Farrar:2003uw,Abdalla:2009mt,Jamil:2009eb,CalderaCabral:2009ja,Micheletti:2009pk,Chen:2008ft,Jackson:2009mz,CalderaCabral:2008bx,He:2008tn,Amendola:2006dg,Costa:2018aoy,Feng:2017usu,Guo:2017deu,Li:2015vla,Li:2013bya,Cai:2017yww}. Although from the fundamental physics viewpoint the mechanism residing behind the interactions between the dark components is yet unknown due to the unknown nature of DM and DE, while simple parametrizations allowing the study of the impact of such interactions at the level of cosmological data exist\footnote{Interestingly, fundamental physics models underlying DM-DE interactions can be found for instance in modified gravity models. In the Einstein frame, many modified gravity models may be written as an effective interacting DM-DE model. To make a concrete example relevant to a model which has been studied considerably in the recent literature, mimetic gravity~\cite{Lim:2010yk,Chamseddine:2013kea,Chamseddine:2014vna,Sebastiani:2016ras} (see also~\cite{Golovnev:2013jxa,Barvinsky:2013mea,Chaichian:2014qba,Deruelle:2014zza,Nojiri:2014zqa,Capela:2014xta,Mirzagholi:2014ifa,Myrzakulov:2015qaa,Arroja:2015wpa,Myrzakulov:2015kda,Cognola:2016gjy,Zheng:2017qfs,Vagnozzi:2017ilo,Casalino:2018tcd,deHaro:2018sqw}) with a non-vanishing potential $V(\phi)$ for the mimetic $\phi$ effectively corresponds to an interacting DM-DE model (see especially the discussion towards the end of Sec.~2 of~\cite{Dutta:2017fjw}, where the fact that this is an interacting DM-DE model is clarified).}. In fact, it has been found that  couplings between DM and DE, as long as sufficiently tiny, are allowed by cosmological data (see e.g. \cite{Salvatelli:2014zta, Eu,Ferreira,Kumar:2016zpg, Yang:2017yme,Yang:2018ubt}).

More interestingly still, such an interaction between the dark components has been shown to be potentially able to reconcile some of the tentative tensions between high- and low-redshift cosmological observables~\cite{Kumar:2016zpg, DiValentino:2017iww}, such as the tension between cosmic microwave background (CMB)~\cite{Ade:2015xua,Aghanim:2016yuo} and local determinations of the Hubble constant $H_0$~\cite{R16}. In addition, an interaction between DM and DE has been shown to be effective in alleviating the tension between the value of $\sigma_8$ inferred from the CMB and the weak lensing measurements~\cite{Pourtsidou_Tram, An}, such as those from the Canada France Hawaii Lensing Survey (CFHTLenS)~\cite{Heymans:2012gg, Erben:2012zw}, the Kilo Degree Survey of$~$450 deg$^2$ of imaging data (KiDS-450)~\cite{Hildebrandt:2016iqg}, and the Dark Energy Survey (DES)~\cite{Abbott:2017wau}. Recently, it was shown that a constraint on the DM-DE scattering cross section of order $\sigma_{\text{DM-DE}} < 10^{-29} \, {\rm cm}^2 \, (m_{\rm DM}/{\rm GeV})$ (where $m_{\rm DM}$ is the DM mass scale) can be obtained by a combination of recent cosmological data~\cite{Kumar_Nunes}. At any rate, it is worth keeping in mind that these tensions are tentative rather than persistent (due to their currently low statistical significance), and might very well be due to systematic effects (see e.g.~\cite{Bernal:2018cxc}).

In the present work, we explore additional features of IDE models, especially related to the stability of the linear perturbations evolution. The presence of an interaction between the dark components changes the dynamics at the perturbative level and consequently the processes leading to structure formation, as well as those determining the properties of the CMB. It has been found that the DE equation of state (EoS), $w_x$, plays a determining role when the structure formation process is considered. In fact, as soon as the DE EoS crosses the phantom divide given by $w_{x}=-1$, instabilities appear due to dark sector perturbations at early times, leading to the blow-up of curvature perturbations on super-horizon scales (see also~\cite{Vikman:2004dc,Easson:2016klq,Deffayet:2010qz} for relevant works). In order to avoid such problems, two separate regions in the parameter space for the DE EoS are usually explored: the non-phantom region where $w_{x} > -1$ and the phantom region where $w_{x} < -1$. Having to examine these two regions of parameter space separately is, of course, undesirable, since it can obscure important cosmological dynamics. Recently, in~\cite{Yang:2017zjs,Yang:2017ccc} it was shown that such problem might be bypassed by considering specific EoS-dependent forms for the coupling between DM and DE. In other words, a choice of some novel forms for the coupling function may be used to construct novel IDE models which are stable for both phantom and non-phantom values of the EoS of DE. In order to avoid divergences at the perturbative level within the context of IDE models, a parametrized post-Friedmann framework has also been applied \cite{Skordis_Pourtsidou_Copeland, Zhang, Guo}.

In this paper, we extend the previous analyses~\cite{Yang:2017zjs,Yang:2017ccc} by choosing coupling functions proportional to the energy density of the DM. In normal circumstances, such couplings are well-known to present instabilities at early times. Motivated by the possibility of determining whether certain couplings offer a more satisfactory description of current observational data, with the restriction that the resulting dark sector's perturbations be stable, we use the method proposed in~\cite{Yang:2017zjs,Yang:2017ccc} and introduce an EoS-dependence in the DM-DE coupling: we obtain in such a way two IDE models with couplings which are still proportional to the DM energy density, but whose dark sector perturbations are now stable. Thus, with the methodology presented here, we are granted the possibility of examining previously problematic couplings presented in the literature.

In our work, we determine novel constraints on the DM-DE coupling parameter for two stable IDE models where the coupling functions (determining the energy flow between the dark components) are proportional to the energy density of the DM particles. Our analysis also indicates a mild preference for the DE EoS being of phantom character ($w_x<-1$) for both IDE models, in some cases with a significance greater than 95\% CL. In addition, we also perform a Bayesian model comparison analysis, by computing the Bayesian evidence for the two IDE models we consider and comparing it to the evidence for the standard $\Lambda$CDM model. The model comparison part of our analysis is an important aspect often not considered in previous related works: in fact, extended models with additional parameters with respect to $\Lambda$CDM (such as our two IDE models, which feature two additional parameters) are generally expected to improve the fit to cosmological data when compared to $\Lambda$CDM. However, only if the improvement in fit is sufficiently large the model can be considered statistically preferred over $\Lambda$CDM, thus, overcoming the fact that the Bayesian evidence penalizes the presence of additional parameters (this can be seen as a quantification of Occam's razor, in other words, models should not be unnecessarily complicated unless deemed so by the data). Our Bayesian model comparison analysis therefore allows us to quantify the statistical preference (if any) for the two IDE models over $\Lambda$CDM.

This work has been organized as follows. In Section~\ref{sec-2}, we discuss the basic equations of IDE models both at background and perturbation level. In Section~\ref{sec-models}, we introduce the two stable parametric interacting DM-DE models, as well as the resulting perturbation equations. In Section~\ref{sec-data} we discuss the observational data used, with results of the analysis presented in Section~\ref{sec-results}. Finally, in Section~\ref{sec-discuss}, we provide concluding remarks. 

\section{Background and perturbations equations}
\label{sec-2} 

We consider a homogeneous and isotropic model for the universe, described by the Friedmann-Lema\^{i}tre-Robertson-Walker (FLRW) metric, whose line element is given by:
\begin{eqnarray}
ds^2 = -dt^2 + a^2 (t) \left[ \frac{dr^2}{1-\kappa r^2} + r^2 \left( d \theta^2 + \sin^2 \theta d \phi^2 \right)  \right]
\end{eqnarray}
where $a(t)$ is the expansion scale factor and $\kappa$ is a constant parametrizing the curvature of space, which can be flat, open or closed for $\kappa=0$, $-1$ and $+1$ respectively. In this work, we shall only consider a spatially flat Universe, and hence fix $\kappa=0$. The matter sector is assumed to be minimally coupled to gravity. The total energy density in the matter sector comes from the contributions of four components, namely, radiation, baryons, cold DM and DE. In the following, we shall use this notation: $\rho_i$ denotes the energy density of the $i$-th species where $i$ runs over radiation ($r$), baryons ($b$), DM ($c$) and DE ($x$). In a similar fashion, we identify $p_i$ as being the pressure of the $i$-th species.

As previously discussed, we assume that the DM and DE fluids interact with each other. The baryon and radiation components are subject to the standard continuity equations. On the other hand, since DM and DE interact among each other, their continuity equations are modified. In particular, in a FLRW universe, these equations are modified to the following:
\begin{eqnarray}
&&\rho_c^\prime = - 3 \mathcal{H} \rho_c - a Q_c,\label{cons-dm}\\
&&\rho_x^\prime = - 3 \mathcal{H} (1+w_x) \rho_x - a Q_x,\label{cons-de}
\end{eqnarray}
where $w_x = p_x/\rho_x$ is the EoS parameter of the DE, the energy transfer rate or the interaction rate is $Q_c = - Q_x = Q$, the prime has been taken with respect to the conformal time, and $\mathcal{H} = a^\prime/a$ is the conformal Hubble parameter. At this point one can consider a variety of phenomenological functional forms for the energy transfer rates between the dark sectors. The predictions for various functional forms, which affect the dynamics of the universe, then can and should be compared against observations. The expansion rate of the universe can be written as:
\begin{eqnarray}\label{Hubble}
\mathcal{H}^2  = \frac{8 \pi G}{3} \, a^2 \sum _{i} \rho_i~.
\end{eqnarray}
Thus, using Eqs.~(\ref{cons-dm},\ref{cons-de}) together with Eq.~(\ref{Hubble}), one can easily determine the background dynamics of IDE models.

At this point, all that remains to choose is the functional form for the DM-DE interaction rate $Q$. Common choices for the interaction rates depend on the energy densities of the DM and DE (e.g. $Q \propto \rho_c$, $Q \propto \rho_x$, or more complicated forms). However, these commonly used choices usually lead to instabilities for certain values of the DE EoS $w_x$, so that this parameter needs to be fixed or restricted to certain regions of parameter space. This undesirable feature leads of course to a loss of information when testing these models against observations. The problem can, nevertheless, be resolved in a class of special interaction models recently proposed~\cite{Yang:2017zjs, Yang:2017ccc}. In this work, we extend the analysis to include additional interaction models which would under normal circumstances feature instabilities. In order to do so, it is necessary to discuss perturbations around the flat FLRW background.

In what follows, we consider the perturbed FLRW metric, whose line element is given by the following:
\begin{eqnarray}
ds^{2}=a^{2}(\tau )\Bigg[-(1+2\phi )d\tau ^{2}+2\partial _{i}Bd\tau dx^{i}+
\Bigl((1-2\psi )\delta _{ij}+2\partial _{i}\partial _{j}E\Bigr)dx^{i}dx^{j}%
\Bigg]\,,
\end{eqnarray}
where $\tau$ denotes conformal time whereas $\phi $, $B$, $\psi $ and $E$ are the  the gauge-dependent scalar perturbation quantities. Covariant conservation of energy-momentum is described by the following equations:
\begin{equation}
\nabla _{\nu }T_{A}^{\mu \nu }=Q_{A}^{\mu },~~~~\sum\limits_{\mathrm{A}}{%
Q_{A}^{\mu }}=0,
\end{equation}%
where the symbol $A$ stands for either DM or DE, and the quantity $Q_{A}^{\mu }$ is defined by :
\begin{eqnarray}
Q_{A}^{\mu }=(Q_{A}+\delta Q_{A})u^{\mu }+a^{-1}(0,\partial
^{i}f_{A}), 
\end{eqnarray}
with $u^{\mu}$ the velocity four-vector and $Q_A$ the background energy transfer, i.e. $Q_A = Q$. We here consider the simplest possibility wherein the momentum transfer potential is  zero in the rest frame of DM, i.e. $k^{2}f_{A}=Q_{A}(\theta -\theta _{c})$. Within the perturbed
FLRW metric, the perturbation equations for the 
fluid $A$ are given by the following~\cite{Majerotto:2009np, Valiviita:2008iv, Clemson:2011an}:
\begin{widetext}
\begin{eqnarray}
\delta_A^{\prime} + 3 \mathcal{H} \left(c_{sA}^2 - w_A \right) \delta_A + 9 \mathcal{H}^2 \left(1+w_A \right) \left(c_{sA}^2- c_{aA}^2 \right)\frac{\theta_A}{k^2} + \left(1+w_A \right) \theta_A -3 \left(1+w_A \right) \psi^{\prime} + \left(1+w_A \right) k^2 \left(B- E^{\prime} \right)\nonumber\\ = \frac{a}{\rho_A} \left(\delta Q_A - Q_A \delta _A \right) + \frac{a Q_A}{\rho_A} \left[\phi + 3 \mathcal{H} \left(c_{sA}^2- c_{aA}^2 \right)\frac{\theta_A}{k^2}\right],\label{density-perturb}\\
\theta_A^{\prime} + \mathcal{H} \left(1-3 c_{sA}^2  \right)\theta_A - \frac{c_{sA}^2}{1+w_A} k^2 \delta_A -k^2 \phi = \frac{a}{(1+w_A)\rho_A} \Bigl[ \left(Q_A \theta -k^2 f_A \right) - \left(1+ c_{sA}^2 \right) Q_A \theta_A \Bigr],\label{velocity-perturb}
\end{eqnarray}
\end{widetext}
where we have neglected shear stress, i.e. $\pi_A = 0$. In addition, $c_{sA}^2$, $c_{aA}^2$,
are respectively the square of the physical and adiabatic sound speed of the fluid $A$, and $\theta = \theta_{\mu}^{\mu}$ is the volume expansion scalar. In order to avoid serious instabilities, $c_{sA}^2 \geq 0$ needs to be assumed in general, as we shall do here.

In relation to the early-time stability of linear perturbations in the dark sector, we now need to concentrate on a second issue, i.e. that of pressure perturbations of the DE. One clearly sees that the behaviour of the DE EoS parameter might lead to a divergence in the limiting case where $w_x \rightarrow -1$, as follows~\cite{Gavela:2009cy}:
\begin{eqnarray}
\delta p_{x}
=c_{sx}^{2}\delta \rho _{x}+3\mathcal{H}\rho
_{x}(1+w_{x})(c_{sx}^{2}-c_{ax}^{2})\left[ 1-\frac{aQ}{3\mathcal{H}\rho
_{x}(1+w_{x})}\right] \frac{\theta _{x}}{k^{2}}~. 
\label{eq:deltap}
\end{eqnarray}
From Eq.~(\ref{eq:deltap}), we see that the quantity of interest is given by the second factor in the square bracket on the right-hand side. In particular, this should be positive in order to ensure the stability of the model. We therefore define the ``doom-factor" $d$ as follows~\cite{Gavela:2009cy}:
\begin{eqnarray}
d \equiv - \frac{aQ}{[3\mathcal{H}\rho_x(1+w_x)]}\,,
\end{eqnarray}
where stability of the IDE model is obtained for $d \leq 0$. Analyzing the sign of the doom factor will therefore determine the conditions under which a given IDE model is stable.

Let us consider some IDE models commonly examined in the literature. Most of the couplings considered in the literature envision an energy transfer rate of the form $Q =  \xi \bar{Q}$ or $Q = \xi H \bar{Q}$, where $\xi$ is a constant coupling determining the strength of the DM-DE interaction. The function $\bar{Q}$ is generally chosen to depend on the energy densities of the dark fluids ($\rho_c$ and $\rho_x$) as well as their first or second order derivatives with respect to some suitable variable. We focus on the choice of functional form of $Q$ where the Hubble factor appears explicitly, i.e. $Q = \xi H \bar{Q}$. In this case, the doom factor is given by the following:

\begin{eqnarray}\label{doom}
d= - \frac{\xi \, \bar{Q}}{3 \rho_x (1+w_x)}.
\end{eqnarray}
We can make a few observations from Eq.~(\ref{doom}). If the function $\bar{Q}$ is chosen in such a way that it is positive (e.g. one may recall the choices $\bar{Q} = \rho_c, \rho_x, (\rho_c +\rho_x)$ which are among the most used and well-known interaction terms in the literature, where obviously the energy densities for DM and DE are positive) the interaction model becomes stable for the following two parameters ranges:
\begin{itemize}
\item $\xi > 0$ and $w_x > -1$, or
\item $\xi < 0$ and $w_x < -1$
\end{itemize}
The scenario with $w_x = -1$ has been included as a limiting case. Allowing $w_x$ to move freely and possibly cross the phantom divide $w_x=-1$ for the above interaction model, will lead to divergences in the doom factor and consequently to instabilities, which render the model unphysical. The discontinuity in the parameter space of $w_x$ inevitably leads to information loss when comparing the model against observations. In the next Section, we will deal with this problem and propose two novel stable IDE models.

\section{Two stable interacting dark energy models}
\label{sec-models}

In this section, let us propose two new couplings in the dark sector which circumvent the problem discussed above. We introduce a factor of the $(1+w_x)$ to the energy transfer rate in order to obtain a model which is stable for any choice of $w_x$, and hence test the entire parameter space of $w_x$ against observations. In what follows, we propose a general parametric form for the energy transfer rate given by the following:
\begin{eqnarray}\label{Q_function1}
Q= (1+w_x) H f(\xi, \rho_c, \rho_{x}),
\end{eqnarray}
with $f(\xi, \rho_c, \rho_{x})$ representing a generic function of the energy densities of the dark components as well as of the coupling parameter $\xi$. Assuming Eq.~(\ref{Q_function1}), the doom factor is given by the following:
\begin{eqnarray}\label{doom_factor}
d= - \frac{f(\xi, \rho_c, \rho_{x})}{3 \rho_x}.
\end{eqnarray}

A brief comment is in order at this point. Although phenomenological in nature, the type of coupling given in Eq.~(\ref{Q_function1}) could actually be quite problematic from a fundamental physics point of view. On the one hand, one would generically expect local interactions between DM and DE particles/fields to determine the interaction rate appearing in Eq.~(\ref{Q_function1}). However, in the given form the latter depends on the Hubble rate $H$, a global quantity, raising the question of how local interactions should ``know" about the global expansion rate. On the other hand, a more serious issue arises if one investigates the quantum field theory of coupled DE models more thoroughly. The fact that the dynamical scales of DM and DE are so different leads to the expectation that the two components should be decoupled, and hence should not interact very efficiently. Insisting on couplings between DM and DE strong enough to lead to observational signatures will then generically lead to large quantum corrections which can spoil the successful behavior of the DE component, while long-range DE-induced forces on the DM component can spoil agreement with observations~\cite{DAmico:2016jbm}. These issues can moreover exacerbate the cosmological constant problem, and are not limited to interacting DM-DE models but are more generically present in models wherein fundamental parameters are varying across cosmological timescales~\cite{Marsh:2016ynw}. Interestingly, one can still envision coupled DE models wherein the above problems are alleviated by considering scenarios featuring multiple axions, some of which would constitute the DM while others still would constitute the DE~\cite{DAmico:2016jbm}.

At any rate, in this paper we shall not be concerned with the microscopical origin of the considered DM-DE interactions. Instead, we will simply take the energy transfer rate given in Eq.~(\ref{Q_function1}) as a convenient phenomenological parametrization. From the historical point of view, in fact, works focusing on observational constraints on coupled DE have often adopted similar coupling functions, while being agnostic as to the fundamental microscopic origin of such couplings and associated problems. One of the reason for such a choice is that DM and DE are conveniently modeled as fluids on cosmological scales, and a mixing between fluids at the level of background evolution such as in Eq.~(\ref{Q_function1}) is conceivable and very simple to model. Keeping these caveats concerning the phenomenological nature of the given coupling in mind, in the following, let us introduce two models.

\subsection{\textit{Model IDErc1}}

It is a well-known fact that the choice $Q \propto \rho_c$ leads to instabilities due the coupling in the dark sector perturbations at early times, with curvature perturbations blowing up on super-horizon scales. Motivated by the possibility of addressing this problem through the $w_x$-dependent term in $Q$, and in order to test our proposal, we propose a first interaction where we make the simple choice of the dark coupling being only a function of the DM energy density, that is, $f = 3 \xi \rho_c$. The factor $3$ is introduced here just for mathematical convenience. Specifically, we consider a coupling function given by:
\begin{eqnarray}\label{model1}
Q =3 (1+w_x) H \xi \rho_c\,.
\end{eqnarray}
In this case, we find that the doom factor is give by $d_{IDErc1}= -\xi \rho_c /\rho_x$. Thus, the stability of the model is ensured by the condition $d\leq0$, whose fulfillment only depends on the sign of $\xi$ and implies $\xi >0$. We refer to this IDE model with coupling given by Eq.~(\ref{model1}) as \textit{\textbf{IDErc1}}.

It is important to note that the sign of the coupling function will depend on the dynamical nature of the DE (that is, on $w_x$) and on the coupling parameter $\xi$. Since $\xi \geq 0$ is a necessary condition to have the stability, the behavior of $w_x$ (phantom or quintessence-like) will determine the nature of the coupling between the dark components. In particular, for phantom DE ($w_x < -1$), the factor $1+w_x$ becomes negative, and hence $Q<0$. This corresponds to a scenario where energy transfer occurs from DE to DM. Similarly, for quintessence-like DE ($w_x>-1$), we have $Q>0$, with energy flowing from DM to DE. Our approach thus opens a whole new perspective towards investigating the interactions between DM and DE, envisioning scenarios where the dynamical nature of DE determines the form of the coupling function. To summarize, to quantify the dark coupling we need to analyze both the parameters $w_x$ and $\xi$. 
Within our approach, the sign of $Q$ is influenced by the DE EoS, and $Q < 0$ is possible for a phantom DE.

Within the \textit{IDErc1} model, the linear perturbation equations for the DE and DM components in synchronous gauge are given by the following:
\begin{eqnarray}
\delta _{x}^{\prime } &=&-(1+w_{x})\left( \theta _{x}+\frac{h^{\prime }}{2}%
\right) -3\mathcal{H}(c_{sx}^{2}-w_{x})\left[ \delta _{x}+3\mathcal{H}%
(1+w_{x})\frac{\theta _{x}}{k^{2}}\right]   \notag \\
&+&3\mathcal{H}\xi (1+w_{x})\frac{\rho _{c}}{\rho _{x}}\left[ -\delta _{x}+\delta
_{c}+\frac{\theta +h^{\prime }/2}{3\mathcal{H}}%
+3\mathcal{H}(c_{sx}^{2}-w_{x})\frac{\theta _{x}}{k^{2}}\right] , \\
\theta _{x}^{\prime } &=&-\mathcal{H}(1-3c_{sx}^{2})\theta _{x}+\frac{%
c_{sx}^{2}}{(1+w_{x})}k^{2}\delta _{x}+3\mathcal{H}\xi \frac{\rho _{c}}{\rho _{x}} \left[ \theta
_{c}-(1+c_{sx}^{2})\theta _{x}\right] , \\
\delta _{c}^{\prime } &=&-\left( \theta _{c}+\frac{h^{\prime }}{2}\right) -3%
\mathcal{H}\xi (1+w_{x})\frac{\theta +h^{\prime }/2}{3\mathcal{H}} , \\
\theta _{c}^{\prime } &=&-\mathcal{H}\theta _{c}.  \label{eq:perturbation1}
\end{eqnarray}

\subsection{\textit{Model IDErc2}}

Let us now consider a second interaction scenario, where the dark coupling instead depends on linear combinations of the interaction rates $Q \propto \rho_c$ and $Q \propto \rho_x$. In particular, we assume $f = 3 \xi (\rho_c+\rho_x)$. Hence, the coupling function for this model is given by the following:
\begin{eqnarray}\label{model2}
Q=3(1+w_x)H\xi(\rho_c+\rho_x).
\end{eqnarray}
We refer to this IDE model as \textit{\textbf{IDErc2}}. For this model, the doom factor is given by $d_{IDErc2}= -\xi (\rho_c + \rho_x)/\rho_x$. As for the \textit{IDErc1} model, we can see that the stability of the model is guaranteed for $\xi > 0$. The arguments concerning the sign of $Q$ (and hence the direction of energy flow between DM and DE) and its relation to the phantom or quintessence-like nature of the DE EoS $w_x$ we presented for the \textit{IDErc1} model hold for this model too.

The linear perturbation equations for the DE and DM components in the \textit{IDErc2} model, in synchronous gauge, are given by:

\begin{eqnarray}
\delta _{x}^{\prime } &=&-(1+w_{x})\left( \theta _{x}+\frac{h^{\prime }}{2}%
\right) -3\mathcal{H}(c_{sx}^{2}-w_{x})\left[ \delta _{x}+3\mathcal{H}%
(1+w_{x})\frac{\theta _{x}}{k^{2}}\right]   \notag \\
&+&3\mathcal{H}\xi (1+w_{x})\frac{\rho _{c}+\rho _{x}}{\rho _{x}}\left[ -\delta _{x}+\frac{\rho_c\delta_c+\rho_x\delta_x}{\rho_c+\rho_x}+\frac{\theta +h^{\prime }/2}{3\mathcal{H}}%
+3\mathcal{H}(c_{sx}^{2}-w_{x})\frac{\theta _{x}}{k^{2}}\right] , \\
\theta _{x}^{\prime } &=&-\mathcal{H}(1-3c_{sx}^{2})\theta _{x}+\frac{%
c_{sx}^{2}}{(1+w_{x})}k^{2}\delta _{x}+3\mathcal{H}\xi \frac{\rho _{c}+\rho _{x}}{\rho _{x}} \left[ \theta_{c}-(1+c_{sx}^{2})\theta _{x}\right] , \\
\delta _{c}^{\prime } &=&-\left( \theta _{c}+\frac{h^{\prime }}{2}\right) +3%
\mathcal{H}\xi (1+w_{x})\frac{\rho _{c}+\rho _{x}}{\rho _{c}}\left(\delta_c-\frac{\rho_c\delta_c+\rho_x\delta_x}{\rho_c+\rho_x}-\frac{\theta +h^{\prime }/2}{3\mathcal{H}}\right) , \\
\theta _{c}^{\prime } &=&-\mathcal{H}\theta _{c}.  \label{eq:perturbation2}
\end{eqnarray}

Notice that when removing the $(1+w_x)$ factor, the two models considered are closely related to those considered in~\cite{Valiviita:2008iv}. The unstable nature of such models when the $(1+w_x)$ factor is not present was studied there and can be seen by inspecting Fig.~1 (lower left panel)  and Fig.~3 (lower panel) of~\cite{Valiviita:2008iv}. There it is clearly shown that, for choices of cosmological parameters which make the models unstable, the gauge-invariant curvature perturbation $\zeta$ blows up on super-Hubble scales to values $\vert \zeta \vert \gg 10^{100}$, in the second case with oscillations with amplitude $\vert \zeta \vert > 10^{300}$.

Having defined the two IDE models we will consider in this work, we can now proceed to comparing their predictions against observational data. In particular, we aim to obtain constraints on the parameters governing the coupling between DM and DE ($\xi$ and $w_x$), in addition to the usual cosmological parameters. In the following Sections, we will describe the observational data against which we constrain the IDE models \textit{IDErc1} and \textit{IDErc2}, as well as the result of our analysis.

\section{Observational data and analysis methodology}
\label{sec-data}

To constrain the chosen IDE models, we use a selection of some of the most recent cosmological datasets, described in the following. We have conservatively chosen not to utilize datasets which could be in tension with each other, and hence we have not used local measurements of the Hubble parameter $H_0$ since we included CMB data, and similar  considerations hold for some recent weak lensing measurements.

\begin{itemize}

\item CMB: We use measurements of CMB temperature and polarization anisotropies, as well as their cross-correlations, from the \textit{Planck} satellite~\cite{Adam:2015rua}. In particular, we use a combination of the high- and low-$\ell$ TT likelihoods (in the overall multipole range $2\leq \ell \leq 2508$), as well as a combination of the high- and low-$\ell$ polarization likelihoods~\cite{Aghanim:2015xee}. This dataset combination is usually referred to as Planck TTTEEE+lowTEB. We analyze these datasets using the publicly available Planck likelihood~\cite{Aghanim:2015xee}, which automatically marginalizes over several nuisance parameters describing uncertainties related to calibration, residual foreground contamination, residual beam-leakage, and so on. 

\item BAO: We consider four distinct Baryon Acoustic Oscillations (BAO) distance measurements. In particular, we use data from (i) the 6dF Galaxy Survey (6dFGS) ($z_{\emph{\emph{eff}}}=0.106$)~\cite{Beutler:2011hx}, (ii) the Main Galaxy Sample of Data Release 7 of the Sloan Digital Sky Survey (SDSS-MGS) ($z_{\emph{\emph{eff}}}=0.15$)~\cite{Ross:2014qpa}, (iii) the CMASS sample of Data Release 12 (DR12) of the Baryon Oscillation Spectroscopic Survey (BOSS) ($z_{\mathrm{eff}}=0.57$)~\cite{Gil-Marin:2015nqa} and finally (iv) the LOWZ sample ($z_{\mathrm{eff}}=0.32$) from the same BOSS data release (DR12)~\cite{Gil-Marin:2015nqa}.

\item JLA: We use the Joint light-curve analysis (JLA) sample of Supernovae Type Ia comprising 740 luminosity distance measurements in the redshift interval $z \in [0.01, 1.30]$~\cite{Betoule:2014frx}.

\item CC: Finally, we consider direct measurements the Hubble parameter from cosmic chronometers (CC). Here, we take the recent compilation 30 measurements of the Hubble parameter data in the redshift interval $0<z<2$~\cite{Moresco:2016mzx}, recompiled after significant improvements in the treatment of the associated systematics (mainly associated to the metallicities of the galaxies used in the analyses). These are model-independent measurements of the expansion history relying on massive and passively evolving early-type galaxies which provide standardizable clocks, whose differential age evolution as a function of redshift provides an estimate of the Hubble parameter at high redshifts.

\end{itemize}

We consider a 8-dimensional parameter space described by the usual 6 parameters of the concordance $\Lambda$CDM model (the baryon and cold dark matter physical energy densities $\Omega_bh^2$ and $\Omega_{c}h^2$, the ratio of the sound horizon at decoupling to the angular diameter distance to last scattering $100 \theta_{MC}$, the optical depth to reionization $\tau$, and the amplitude and tilt of the primordial power spectrum of scalar fluctuations $A_s$ and $n_s$), plus two additional parameters: the dark energy equation of state $w_x$, and the strength of the DM-DE coupling $\xi$. We impose flat priors on the 8 parameters, with prior ranges given in Table~\ref{tab_priors}.

\begin{table}[tbp]
\begin{center}
\begin{tabular}{c|c c}
Parameter & Prior (IDE)   \\ \hline
$\Omega_{b} h^2$ & $[0.005,0.1]$   \\
$\tau$ & $[0.01,0.8]$   \\
$n_s$ & $[0.5, 1.5]$  \\
$\log[10^{10}A_{s}]$ & $[2.4,4]$   \\
$100\theta_{MC}$ & $[0.5,10]$   \\
$w_x$ & $[-2, 0]$   \\
$\xi$ & $[0, 2]$ 
\end{tabular}%
\end{center}
\caption{Flat priors on the various cosmological parameters associated to the interacting dark energy models in this work.}
\label{tab_priors}
\end{table}

To sample the posterior distribution of the cosmological parameters, we use the Monte Carlo Markov Chain (MCMC) sampler \texttt{CosmoMC}~\cite{Lewis:2002ah}, which implements an efficient sampling method. We assess the convergence of the generated MCMC chains through the Gelman-Rubin statistic $R-1$~\cite{Gelman-Rubin}.

In addition to a parameter estimation analysis where we infer the cosmological parameters within the two IDE models in light of data, another part of our work consists in a Bayesian model comparison analysis. Such an analysis is typically (although not always) missing in previous similar studies, which typically only assess the improvement in fit brought about by introducing the DM-DE interaction by computing the difference between $\chi^2$ goodness-of-fit, $\Delta \chi^2$, between a given IDE model and a baseline model, usually $\Lambda$CDM. However, the $\Delta \chi^2$ is actually not a good measure of the \textit{statistical preference} for a model over a baseline model (let us for definiteness hereafter take the baseline model to be $\Lambda$CDM), especially if the two models are nested (i.e. the baseline model is recovered as a particular case of the extended model: for instance, in our case the $\Lambda$CDM model is recovered from the two IDE models when $\xi=0$ and $w_x=-1$). In the latter case, an improvement in fit (i.e. a lower $\chi^2$) is \textit{guaranteed} by construction in the extended model, since the fit can in the worst case only be as bad/good as in the baseline model. However, only a large enough increase in the improvement of fit can justify the presence of additional parameters and hence the increased model complexity, in the spirit of Occam's razor.

A Bayesian evidence (BE) calculation formally quantifies the previous statement, by computing the Bayes factor for the extended model over the baseline one. In fact, the BE trades the higher likelihood of the extended model against the increase in model complexity and hence prior volume. It is worth reviewing some basics of BE calculation. Let us consider a dataset $\mathbf{x}$, and two competing models ${\cal M}_i$ and ${\cal M}_j$, described by the parameters $\boldsymbol{\theta}_i$ and $\boldsymbol{\theta}_j$ respectively (where $\boldsymbol{\theta}_i$ can be a subset of $\boldsymbol{\theta}_j$, or vice versa, in the case of nested models). The Bayes factor $B_{ij}$ of model ${\cal M}_i$ with respect to model ${\cal M}_j$, assuming equal prior probabilities for the two models (which is usually the case) is then given by:
\begin{eqnarray}
B_{ij} = \frac{p({\cal M}_i \vert \mathbf{x})}{p({\cal M}_j \vert \mathbf{x})} \,,
\end{eqnarray}
where the $p({\cal M}_i \vert \mathbf{x})$ is the Bayesian evidence of model ${\cal M}_i$ and is given by:
\begin{eqnarray}
p({\cal M}_i \vert \mathbf{x}) = \int d\boldsymbol{\theta}_i\, \pi(\boldsymbol{\theta}_i \vert {\cal M}_i) {\cal L}(\mathbf{x} \vert \boldsymbol{\theta}_i,{\cal M}_i)\,,
\end{eqnarray}
with $\pi(\boldsymbol{\theta}_i \vert {\cal M}_i)$ and ${\cal L}(\mathbf{x} \vert \boldsymbol{\theta}_i,{\cal M}_i)$ the prior on the parameters $\boldsymbol{\theta}_i$ and the likelihood of the data given the model parameters $\boldsymbol{\theta}_i$ respectively (and of course, analogously for the Bayesian evidence of model ${\cal M}_j$). A value of $B_{ij}>1$ indicates that data support the model ${\cal M}_i$ more strongly than model ${\cal M}_j$, which might be the case even when the goodness-of-fit of model ${\cal M}_j$ improves over that of ${\cal M}_i$, but the former is penalized by the increased model complexity.

There exist scales which allow for qualitative interpretations of different values of $B_{ij}$ (or alternatively $\ln B_{ij}$). In this work, we will adopt an alternative to the widely used Jeffreys scale, provided by Kass and Raftery~\cite{Kass:1995loi}, and summarized in Tab.~\ref{tab:jeffreys}.

Computing the BE evidence, and hence the Bayes factor, is notoriously computationally expensive. Recently, progress on the matter was made in~\cite{Heavens:2017hkr,Heavens:2017afc}, where it was noted that the BE can be directly computed from the MCMC chains used to perform parameter estimation. The proposed algorithm, implemented in the \texttt{MCEvidence} code\footnote{The code is publicly available at~\href{https://github.com/yabebalFantaye/MCEvidence}{github.com/yabebalFantaye/MCEvidence}.}, uses $k$th nearest neighbor distances with distances computed using the Mahalanobis distance, where the inverse covariance matrix estimated from the MCMC chains defines the metric, to estimate the Bayesian evidence from the MCMC samples provided by the chains. In this work, we use the \texttt{MCEvidence} code to compute the logarithm of the Bayes factor of the two IDE models against the standard $\Lambda$CDM model, i.e. we calculate $\ln B_{ij}$ (where $i=$\textit{IDErc1} or \textit{IDErc2} and $j=\Lambda$CDM), and then use the revised Jeffreys scale summarized in Tab.~\ref{tab:jeffreys} to quantify the obtained values of $\ln B_{ij}$ in terms of strength of evidence for $\Lambda$CDM (since, as we shall see, the concordance $\Lambda$CDM model appears statistically preferred over the two IDE models).

\begingroup                                                                                                                     
\begin{center}                                                                                                                  
\begin{table}[!h]                                                                                                                
\begin{tabular}{cc}                                                                                                            
\hline\hline                                                                                                                    
$\ln B_{ij}$ & Strength of evidence for model ${\cal M}_i$ \\ \hline
$0 \leq \ln B_{ij} < 1$ & Weak \\
$1 \leq \ln B_{ij} < 3$ & Definite/Positive \\
$3 \leq \ln B_{ij} < 5$ & Strong \\
$\ln B_{ij} \geq 5$ & Very strong \\
\hline\hline                                                                                                                    
\end{tabular}                                                                                                                   
\caption{Revised Jeffreys scale used in this work to qualify the obtained values of the logarithm of the Bayes factor of model ${\cal M}_i$ with respect to model ${\cal M}_j$, $\ln B_{ij}$, in terms of strength of the evidence for model ${\cal M}_i$.}\label{tab:jeffreys}                                                                                                   
\end{table}                                                                                                                     
\end{center}                                                                                                                    
\endgroup 

\section{Results and analysis}
\label{sec-results}

Recall that, by studying pressure perturbations of the IDE fluid, we found that the prefactor $(1+w_x)$ in the coupling function $Q$, appearing in Eq.~(\ref{model1}) for the \textit{IDErc1} model and in Eq.~(\ref{model2}) for the \textit{IDErc2} model, allows us to explore the entire parameter space of $w_x$, bypassing the problematic point $w_x = -1$. Stability in the evolution of linear perturbations for both the \textit{IDErc1} and \textit{IDErc2} models is realized for $\xi >0$. Having ensured their stability, we can now describe the main observational constraints on the parameters of the two IDE models we consider. We constrain the parameters of the models by considering the following three dataset combinations: CMB alone, CMB+BAO, and CMB+BAO+JLA+CC (recall that these datasets are described in detail in Sec.~\ref{sec-data}).

\begin{figure}[!hbt]
\includegraphics[width=0.8\textwidth]{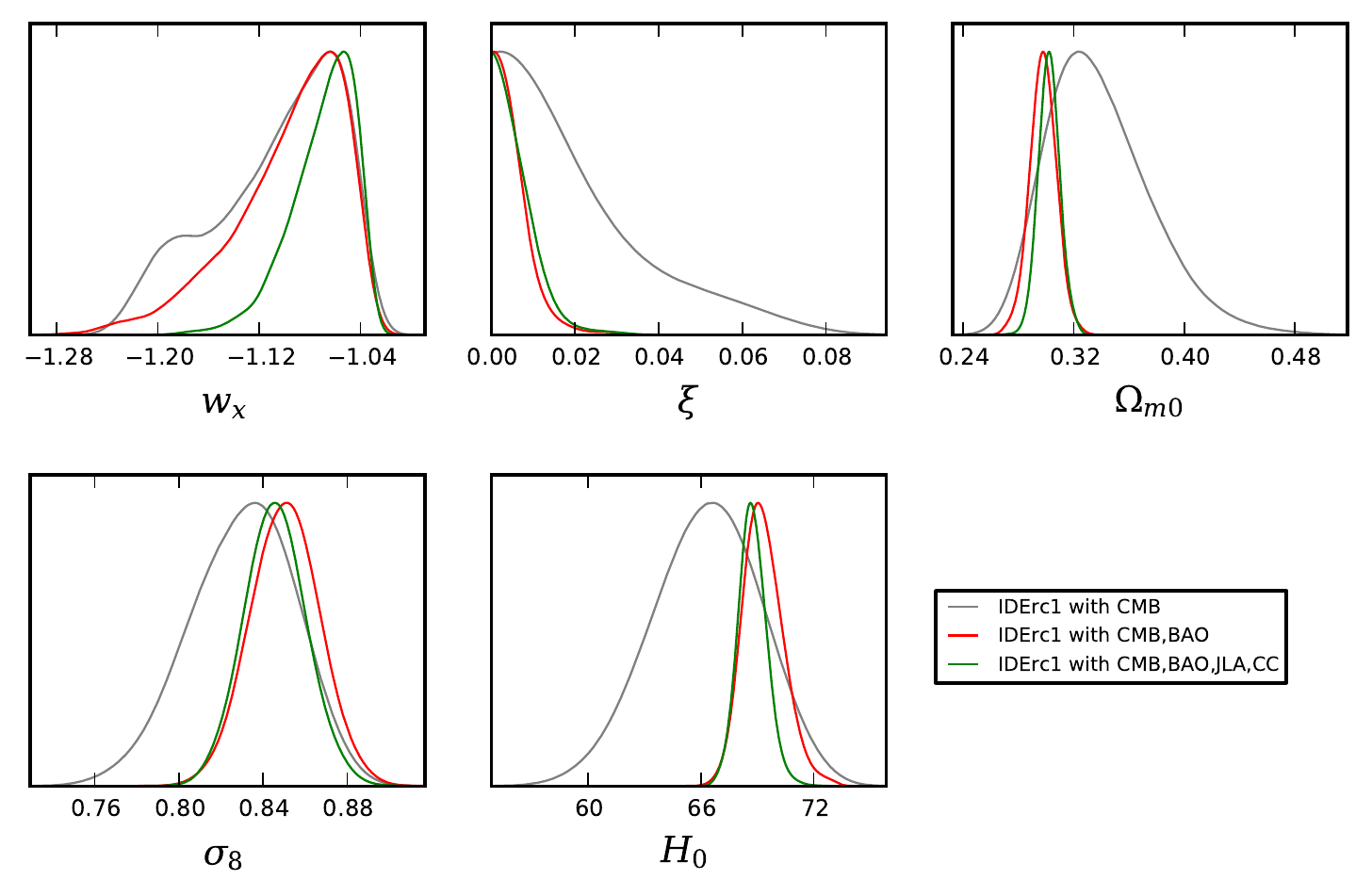}
\caption{1D marginalized posterior distributions for a selection of cosmological parameters of the \textit{IDErc1} model, plotted in arbitrary units and normalized to the maximum of the posterior. The posteriors plotted are obtained from the CMB-only dataset (gray curve), the CMB+BAO dataset combination (red curve), and the CMB+BAO+JLA+CC dataset combination (green curve).}
\label{1D_1}
\end{figure}

\begin{figure}[!hbt]
\includegraphics[width=0.8\textwidth]{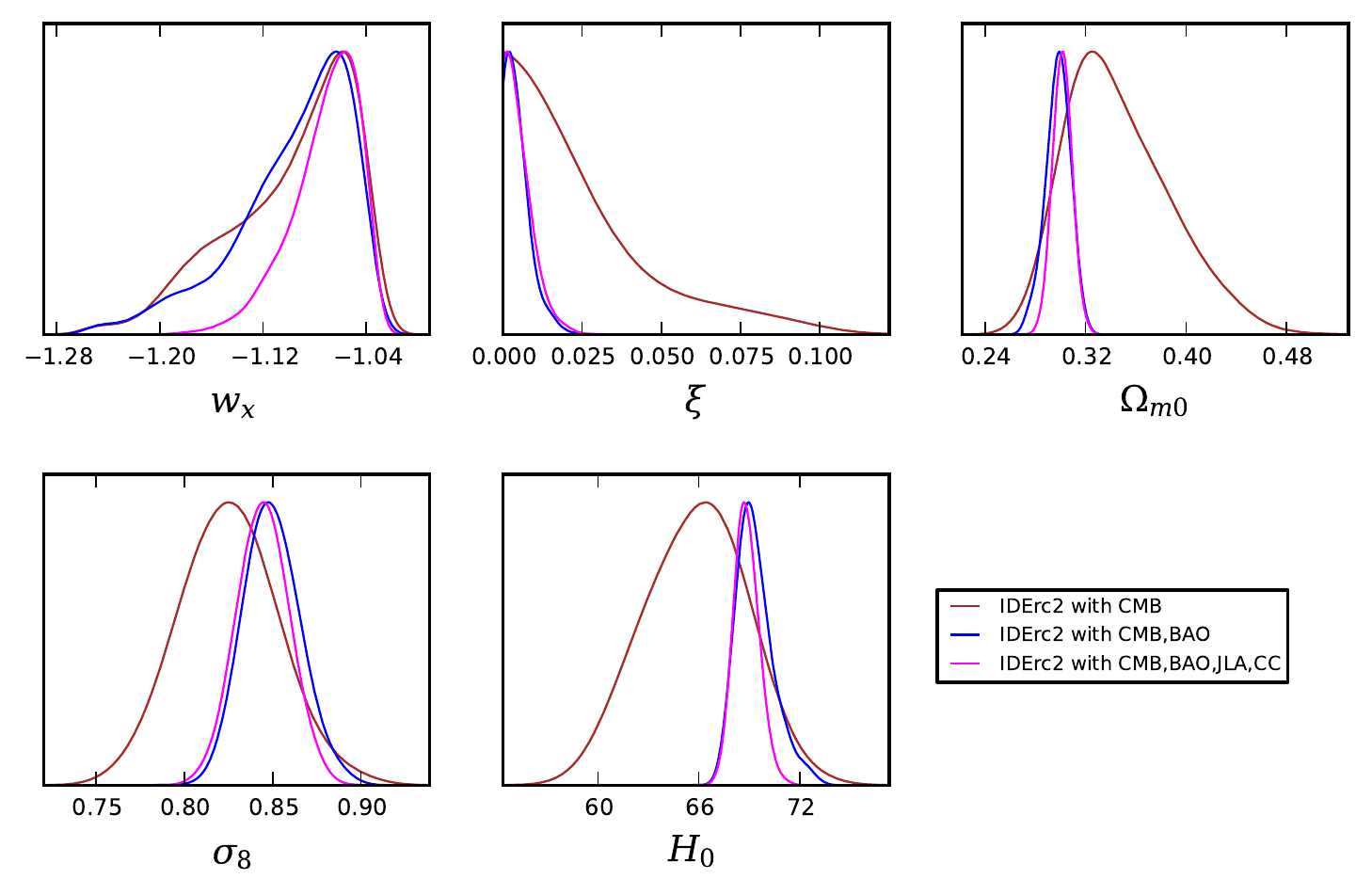}
\caption{As in Fig.~\ref{1D_1} but for the \textit{IDErc2} model.}
\label{1D_2}
\end{figure}

In Tab.~\ref{tab:results_1} and Tab.~\ref{tab:results_2} we summarize our main results. We report the 68\% confidence level (C.L.) credible regions of the cosmological parameters, except for the case of the coupling strength parameter $\xi$, for which we don't have a detection but only an upper limit: in this case, we report the 95\%~C.L. upper limit. In Fig.~\ref{1D_1} and Fig.~\ref{1D_2} we show the 1-dimensional marginalized posterior distributions for selected cosmological parameters within the \textit{IDErc1} and \textit{IDErc2} models respectively. The first thing to notice is that the posterior distribution obtained from the CMB only for $w_x$ is slightly multimodal, with a second smaller peak more pronounced for the \textit{IDErc1} with respect to the \textit{IDErc2}. This second peak seems to prefer a more phantom dark energy equation of state, and disappears as soon as external datasets are added. This quasi-bimodal distribution prevent the CMB only cases to be fully converged, but the impasses is passed when the BAO data select one of the two peaks.

\begin{figure*}[!hbt]
\includegraphics[width=0.8\textwidth]{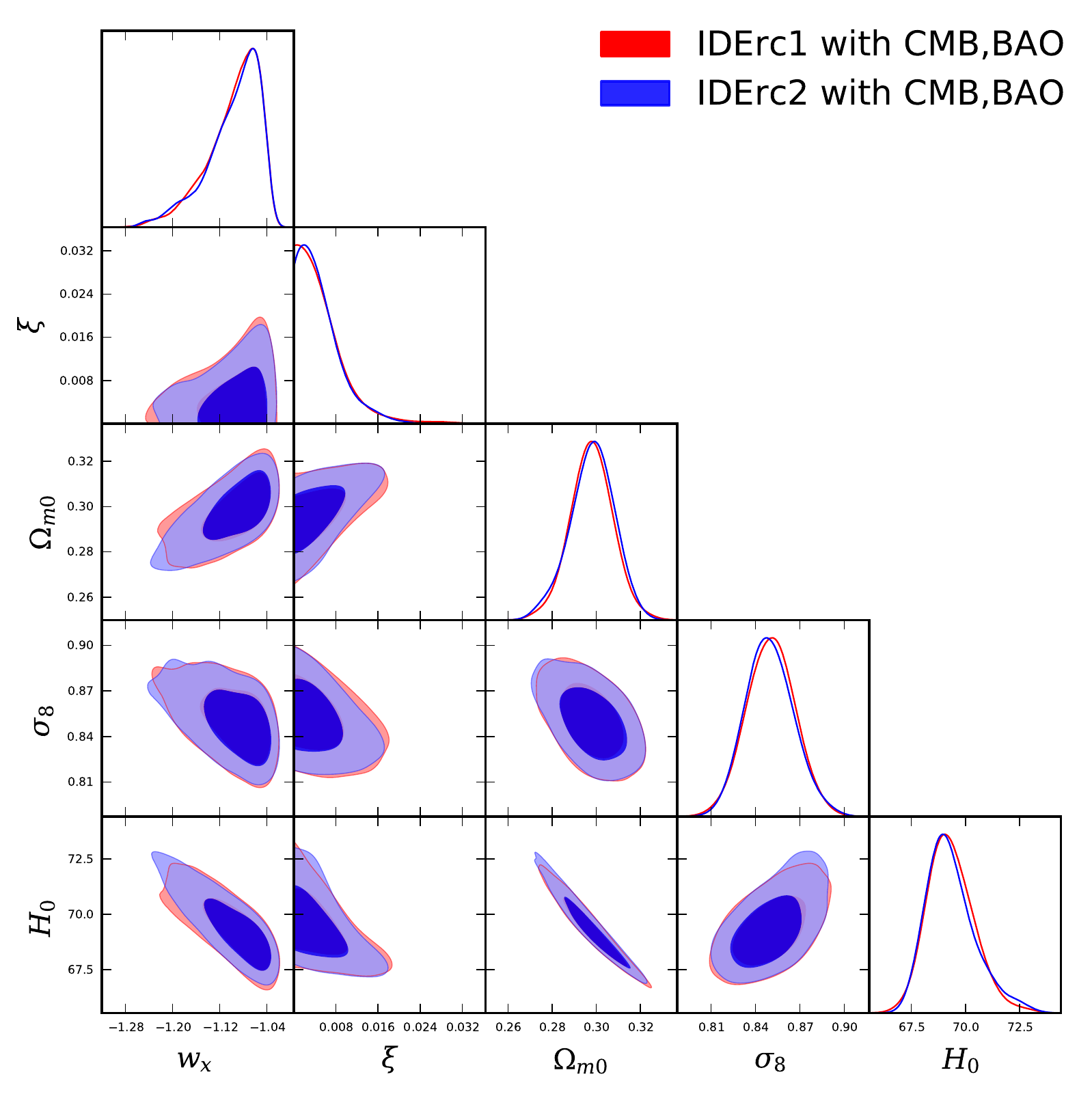}
\caption{Triangular plot showing the 1D marginalized posterior distributions (along the diagonal), and 2D joint posterior distributions for some selected parameters of the \textit{IDErc1} and \textit{IDErc2} models, obtained from the CMB+BAO dataset combination. }
\label{2D_BAO_1}
\end{figure*}

\begin{figure*}[!hbt]
\includegraphics[width=0.8\textwidth]{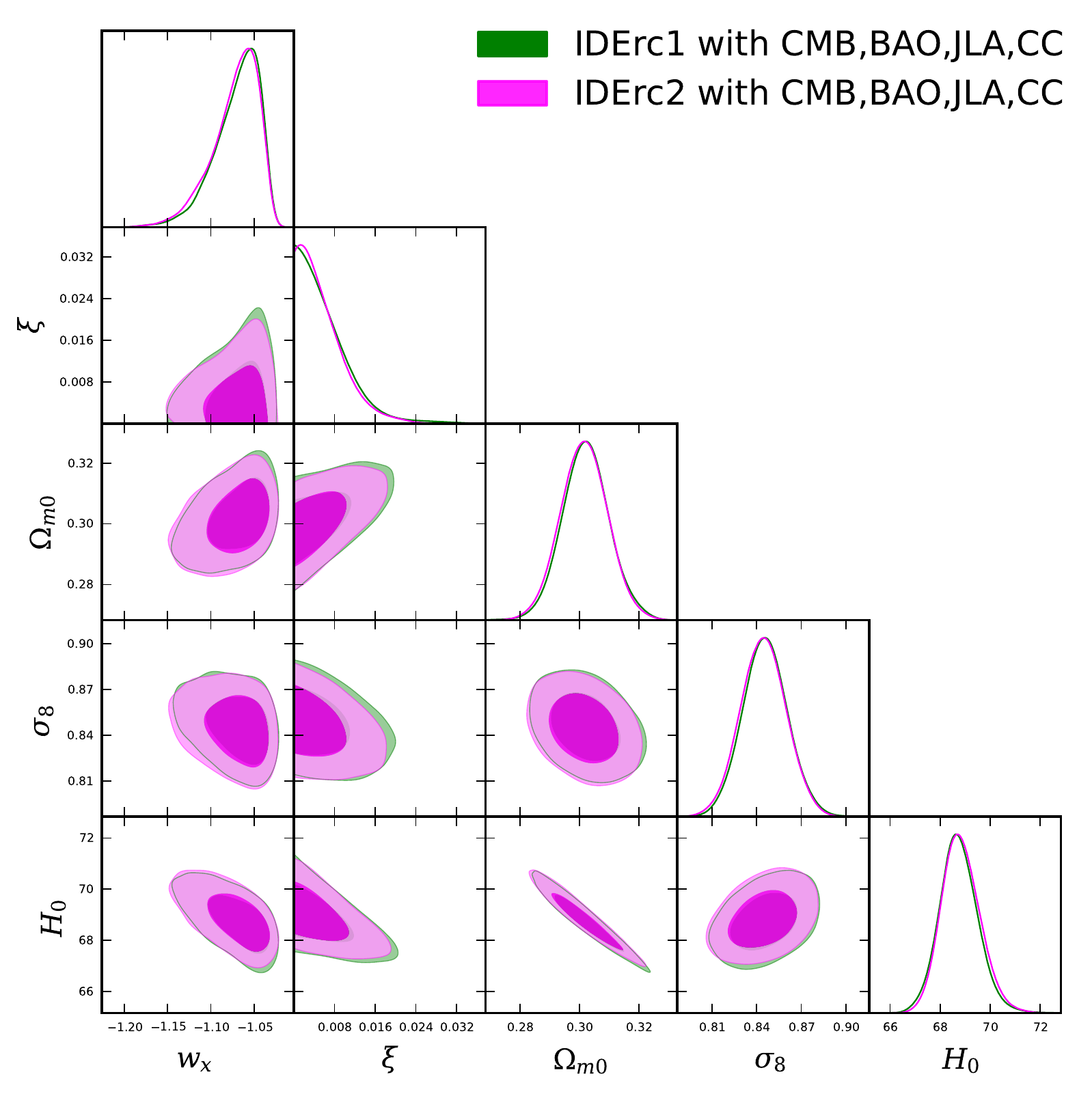}
\caption{As in Fig. \ref{2D_BAO_1} but for the CMB+BAO+JLA+CC dataset combination.}
\label{2D_BAO_2}
\end{figure*}

In Figs.~\ref{2D_BAO_1} and \ref{2D_BAO_2}, we compare the 2D joint and 1D marginalized posteriors of the same parameters among the \textit{IDErc1} and \textit{IDErc2} models, assuming the CMB+BAO, and CMB+BAO+JLA+CC dataset combination, respectively. We can note that the inferred values for the cosmological parameters (including $\xi$) vary very little across the two models. This brings us to conclude that, from the observational point of view, the two models are practically indistinguishable.

It is interesting to note that all the three dataset combinations hint to a phantom behaviour for the EoS of DE ($w_x < -1$) at more than 68\% C.L. for both models. The hint for phantom DE in particular becomes stronger when the CMB+BAO+JLA+CC dataset combination is considered, as the inferred value of $w_x$ is phantom at more than 95\%~C.L.. As discussed previously, the forms of the DM-DE coupling [Eqs.~(\ref{model1},\ref{model2})] in combination with the phantom nature of the DE imply an effective coupling which leads to energy flow from  DE to DM.

\begingroup                                                                                                                     
\squeezetable                                                                                                                   
\begin{center}                                                                                                                  
\begin{table}                                                                                                                   
\begin{tabular}{cccc}                                                                                                        
\hline\hline                                                                                                                    
Parameters & CMB & CMB+BAO & CMB+BAO+JLA+CC \\ \hline
$\Omega_c h^2$ & $    0.1248_{-    0.0044}^{+    0.0027}$ & $0.1201_{-    0.0013}^{+    0.0012}$& $0.1198\pm  0.0011$\\
$\Omega_b h^2$ & $    0.02229\pm 0.00017$ & $    0.02228\pm 0.00015$& $    0.02228_{-0.00016}^{+0.00014}$\\
$100\theta_{MC}$ & $    1.04010_{-    0.00042}^{+    0.00052}$ & $    1.04053_{-    0.00033}^{+    0.00035}$& $    1.04058\pm    0.00031$\\
$\tau$ & $    0.071_{-    0.017}^{+    0.019}$ & $    0.080\pm  0.017$& $    0.082\pm  0.017$\\
$n_s$ & $    0.9682_{-    0.0046}^{+    0.0047}$ & $    0.9736_{-    0.0038}^{+    0.0040}$& $    0.9739\pm  0.0038$\\
${\rm{ln}}(10^{10} A_s)$ & $    3.081_{-    0.033}^{+    0.037}$ & $    3.101_{-    0.033}^{+    0.034}$& $    3.104_{-    0.033}^{+    0.034}$\\
$w_x$ & $   -1.104_{-    0.033}^{+    0.071}$ & $   -1.096_{-    0.026}^{+    0.057}$& $   -1.070_{-    0.016}^{+    0.034}$\\
$\xi$ & $    <0.060$ & $    <0.014$ & $    <0.016$\\
$\Omega_{m0}$ & $    0.339_{-    0.047}^{+    0.030}$ & $    0.2979_{-    0.0094}^{+    0.0095}$& $    0.3022\pm  0.0077$\\
$\sigma_8$ & $    0.830_{-    0.025}^{+    0.028}$ & $    0.850_{-    0.016}^{+    0.017}$& $    0.846\pm 0.015$\\
$H_0[\rm{Km \, s^{-1} \, Mpc^{-1}}]$ & $   66.2_{-    2.9}^{+    3.2}$ & $   69.3_{-    1.2}^{+    1.0}$& $   68.76_{-    0.80}^{+    0.72}$\\
\hline\hline                                                                                                                    
\end{tabular}                                                                                                                   
\caption{Summary of the observational constraints on cosmological parameters (including derived parameters) within the interacting dark energy model \textit{IDErc1}. The constraints are obtained adopting the CMB-only Planck TTTEEE+lowTEB dataset (second column from the left), the CMB+BAO dataset (third column from the left), and the CMB+BAO+JLA+CC dataset (right-most column), see Sec.~\ref{sec-data} for a comprehensive description of these dataset. We report 68\%~C.L. credible intervals for the cosmological parameters, with the exception of the coupling strength $\xi$, for which we only report the 95\%~C.L. Here, $\Omega_{m0}$ denotes the present value of the matter density parameter $\Omega_{m}=\Omega_c+\Omega_b$.}\label{tab:results_1}                                                                                                   
\end{table}                                                                                                                     
\end{center}                                                                                                                    
\endgroup

\begingroup                                                                                                                     
\squeezetable                                                                                                                   
\begin{center}                                                                                                                  
\begin{table}                                                                                                                   
\begin{tabular}{cccc}                                                                                                       
\hline\hline                                                                                                                    
Parameters & CMB & CMB+BAO & CMB+BAO+JLA+CC \\ \hline
$\Omega_c h^2$ & $    0.1255_{-    0.0049}^{+    0.0032}$ & $0.1203\pm 0.0013$& $0.1199\pm  0.0012$\\
$\Omega_b h^2$ & $    0.02229_{-0.00020}^{+ 0.00017}$ & $    0.02228_{-0.00017}^{+0.00015}$& $    0.02228_{-0.00016}^{+0.00015}$\\
$100\theta_{MC}$ & $    1.04007_{-    0.00044}^{+    0.00051}$ & $    1.04055_{-    0.00033}^{+    0.00032}$& $    1.04057\pm    0.00031$\\
$\tau$ & $    0.072\pm 0.018$ & $    0.080\pm  0.017$& $    0.080_{-0.018}^{  +0.017}$\\
$n_s$ & $    0.9680\pm 0.0051$ & $    0.9734_{-    0.0038}^{+    0.0039}$& $    0.9738_{-0.0040}^{+ 0.0037}$\\
${\rm{ln}}(10^{10} A_s)$ & $    3.085_{-    0.035}^{+    0.039}$ & $    3.100_{-    0.032}^{+    0.033}$& $    3.101_{-    0.033}^{+    0.034}$\\
$w_x$ & $   -1.097_{-    0.030}^{+    0.067}$ & $   -1.096_{-    0.027}^{+    0.059}$& $   -1.072_{-    0.016}^{+    0.034}$\\
$\xi$ & $    <0.076$ & $    <0.013$ & $    <0.015$\\
$\Omega_{m0}$ & $    0.346_{-    0.053}^{+    0.033}$ & $    0.298_{-    0.010}^{+    0.011}$& $    0.3016_{-0.0079}^{+0.0078}$\\
$\sigma_8$ & $    0.825_{-    0.029}^{+    0.028}$ & $    0.850_{-    0.018}^{+    0.016}$& $    0.845\pm 0.015$\\
$H_0[\rm{Km \, s^{-1} \, Mpc^{-1}}]$ & $   65.8_{-    3.2}^{+    3.4}$ & $   69.3_{-    1.4}^{+    0.9}$& $   68.84_{-    0.84}^{+    0.70}$\\
\hline\hline                                                                                                                    
\end{tabular}                                                                                                                   
\caption{As in Tab.~\ref{tab:results_1} but for the \textit{IDErc2} model. }\label{tab:results_2}                                                                                                   
\end{table}                                                                                                                     
\end{center}                                                                                                                    
\endgroup 


\begingroup                                                                                                                     
\squeezetable                                                                                                                   
\begin{center}                                                                                                                  
\begin{table}                                                                                                                   
\begin{tabular}{ccc}                                                                                                            
\hline\hline                                                                                                                    
Parameters & CMB & CMB\\ \hline
$\Omega_c h^2$ & $    0.1198\pm 0.0015$ & $0.1196\pm 0.0015$\\
$\Omega_b h^2$ & $    0.02225\pm 0.00016$ & $    0.02229\pm0.00016$\\
$100\theta_{MC}$ & $    1.04077\pm   0.00032$ & $    1.04080\pm0.00031$\\
$\tau$ & $    0.079\pm 0.017$ & $    0.075\pm  0.017$\\
$n_s$ & $    0.9645\pm 0.0049$ & $    0.9649\pm 0.0048$\\
${\rm{ln}}(10^{10} A_s)$ & $    3.094\pm   0.034$ & $    3.085\pm 0.033$\\
$w$ & $   [-1]$ & $   -1.55_{-    0.38}^{+    0.19}$\\
$\Omega_{m0}$ & $    0.3156\pm   0.0091$ & $    0.203_{-    0.065}^{+    0.022}$\\
$\sigma_8$ & $    0.831\pm 0.013$ & $    0.98_{-    0.06}^{+    0.10}$\\
$H_0[\rm{Km \, s^{-1} \, Mpc^{-1}}]$ & $   67.27\pm 0.66$ & $   >81.3$\\
\hline\hline                                                                                                                    
\end{tabular}                                                                                                                   
\caption{Planck TTTEEE+lowTEB constraints assuming the $\Lambda$CDM model (second column) and the $w$CDM model (where the 6 parameters of $\Lambda$CDM plus the dark energy equation of state $w_x$ are varied, third column). Mean values of the parameters are displayed at 68\% C.L.}\label{tab:results_lcdm}                                                                           \end{table}                                                                                                                     
\end{center}                                                                                                                    
\endgroup 

Concerning the coupling parameter $\xi$, we find new upper limits. In particular, for the CMB+BAO+JLA+CC dataset combination, we find $\xi < 0.016$ and $\xi < 0.015$ for \textit{IDErc1} and \textit{IDErc2} respectively, both being 95\%~C.L. upper limits. The other upper limits on $\xi$ obtained from considering other dataset combinations are shown in Tab.~\ref{tab:results_1} and Tab.~\ref{tab:results_2}. It is worth comparing our results to other similar results obtained in the literature. In~\cite{Kumar:2016zpg}, an IDE model with $Q \propto \rho_c$ was investigated, with the coupling parameter found to be $\xi = 0.0008^{+0.0011}_{-0.0011}$, and the equation of state correspondingly $w_x = -0.973^{+0.039}_{-0.039}$, both being 68\%~C.L. credible intervals obtained from analyzing the CMB+BAO+JLA dataset. These constraints are obtained assuming that the energy flow is parallel to the four-velocity of the DM and there is no momentum transfer in the rest frame of DM. It thus follows that the DM velocity perturbations are not affected by the interaction, which can be considered a particular case of the physical situation realized in our scenario. Other constraints on interactions of the form $Q \propto \rho_c$ can be found in the literature, see e.g.~\cite{Yang:2017yme, Nunes_Barboza, Costa_Barboza_Alcaniz, Yan_Wang_Meng}.

\begin{figure*}[!h]
\includegraphics[width=8 cm]{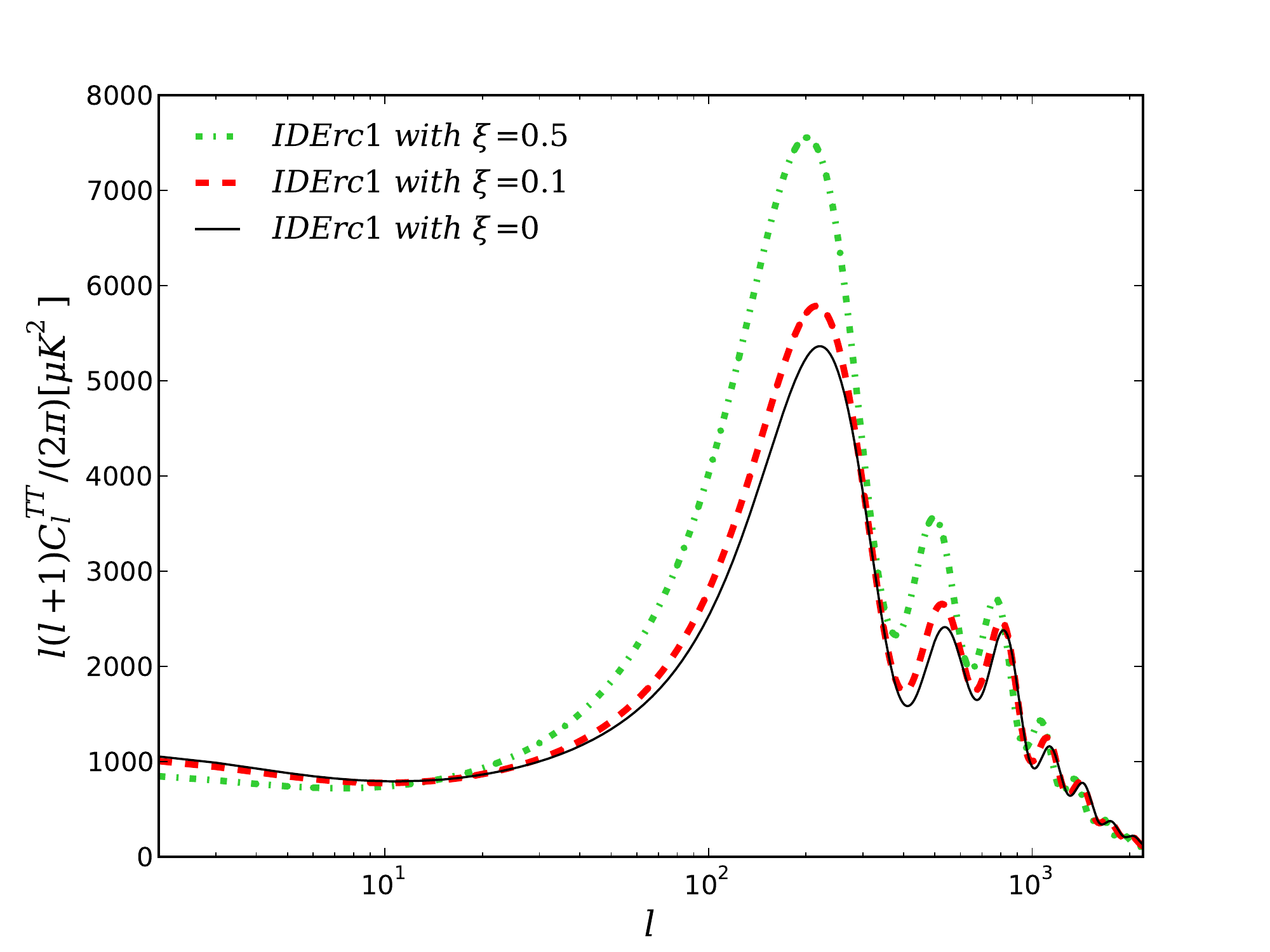} \quad \quad 
\includegraphics[width=8 cm]{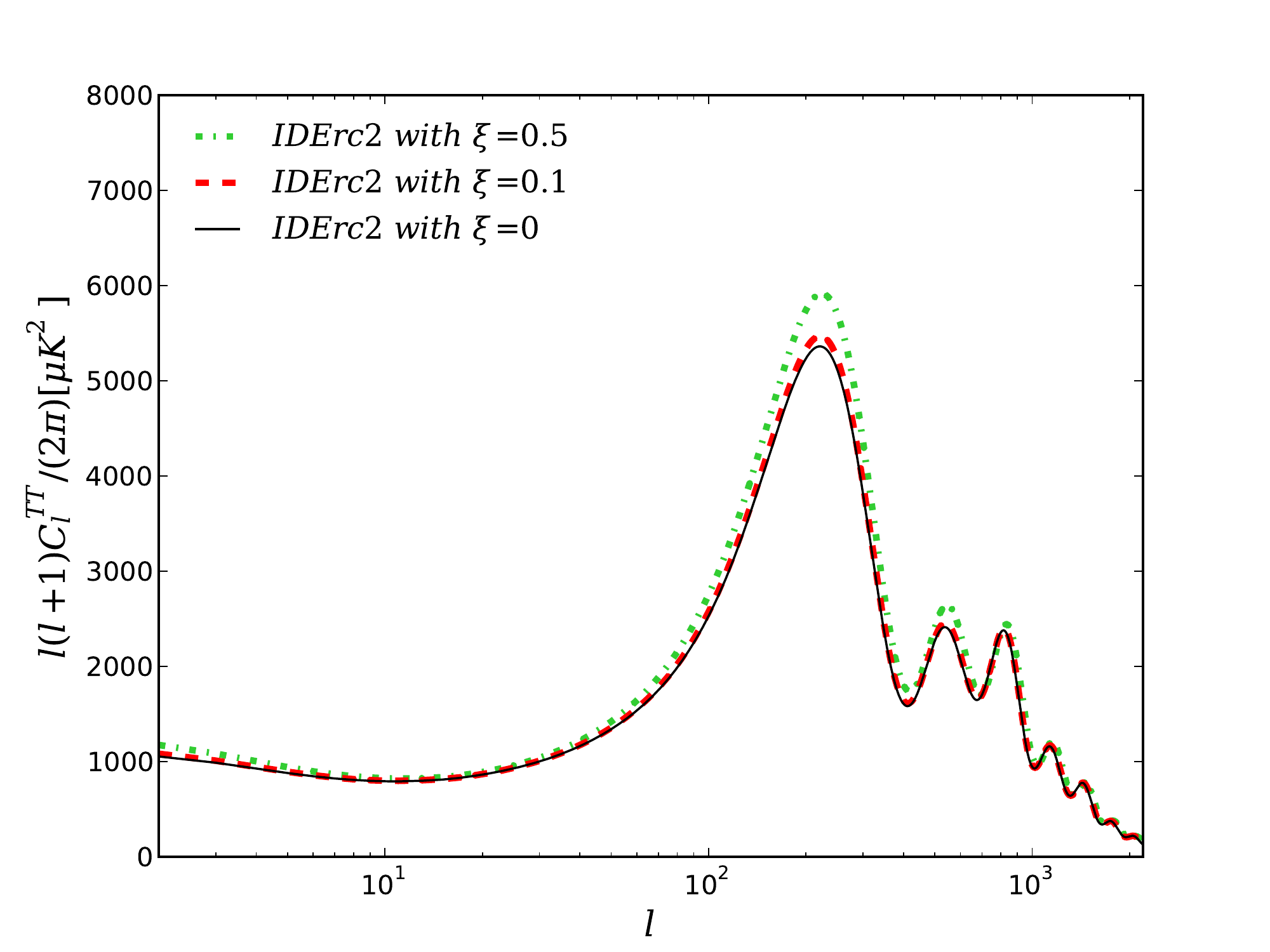}
\caption{Theoretical prediction for CBM TT power spectrum in presence of the DM-DE coupling for different values of $\xi$. Left panel (right panel) shows the effects for the scenarios \textit{IDErc1} (\textit{IDErc2}), respectively.}
\label{cltt}
\end{figure*}

It is worth investigating in more detail the physical effect of the DM-DE interaction we introduced on our cosmological observables, and in particular on the CMB temperature power spectrum $C_{\ell}^{TT}$, since this observable carries a huge amount of information. In Fig.~\ref{cltt}, we plot the theoretical values for $C_{\ell}^{TT}$ as $\xi$ is increased (obviously considering only positive values of $\xi$, as required by the stability of the models), whereas the other parameters are fixed to their mean values determined by their posterior distributions we obtain by analyzing the CMB+BAO+JLA+CC dataset. In particular, for both models we fix $w_x=-1.07$, which implies that the energy flow occurs from DE to DM, as discussed previously. The values at which the other parameters are fixed can be found by examining the right-most column of Tab.~(\ref{tab:results_1}) for model \textit{IDErc1} and Tab.~(\ref{tab:results_2}) for model \textit{IDErc2}. In order to be pedagogical and boost the impact of increasing $\xi$ on the temperature power spectrum, we have considered quite extreme values of $\xi$, namely $\xi=0.1$ and $\xi=0.5$, in addition to the case $\xi=0$ where there is no interaction between DM and DE. While these are unphysically large values, they help us assessing the impact of the DM-DE interaction.

The first thing we notice from Fig.~\ref{cltt} is that the impact of increasing $\xi$ is more pronounced for the \textit{IDErc1} model than for the \textit{IDErc2} model. Let us therefore focus on the \textit{IDErc1} model (left panel) for definiteness. From Fig.~\ref{cltt}, left panel, we see that the main effect of increasing $\xi$ is a boost in the amplitude of all the acoustic peaks (especially the first one), as well as a shift of all peaks to lower multipoles $\ell$, i.e. to larger angular scales, when compared to the case of no interaction ($\xi= 0$). Finally, we also see from the low-$\ell$ tail that as $\xi$ is increased, there is a decrease in the amplitude of the late integrated Sachs-Wolfe effect for the \textit{IDErc1} model, and an increase in the amplitude of the same effect for the \textit{IDErc2} model. However, the effects are very small for realistic values of $\xi$ and masked by the large cosmic variance in the corresponding multipole region.

The shift of all peaks to lower multipoles effect is related to the well-known geometrical degeneracy present with CMB data, and hence we expect $\xi$ to be strongly degenerate with parameters controlling this degeneracy, such as $\Omega_m$ and $H_0$. In particular, we expect $\xi$ to be positively correlated with $\Omega_m$ and negatively correlated with $H_0$. This expectation is in fact borne out by examining the joint posteriors of $\xi$-$\Omega_m$ and $\xi$-$H_0$ in Fig.~\ref{2D_BAO_1} and Fig.~\ref{2D_BAO_2}. From Fig.~\ref{cltt} we see that there is little room for substantial variations of $\xi$ within the error bars of the measurements, hence driving the strong upper limits we obtain on $\xi$.

As discussed in Section~\ref{sec-models}, the sign of $Q$, which quantifies the direction of the energy flow between DM and DE, depends directly on $w_x$, since $H > 0$ and $\rho_{i} > 0$, with $i= c, \, x$. Let us define the effective coupling parameter $\xi_{\rm eff}$ as $\xi_{\rm eff}=(1+w_x)\xi$. For the \textit{IDErc1} model we have furthermore that $\xi_{\rm eff}=Q/3H\rho_c$, while for the \textit{IDErc2} model we have that $\xi_{\rm eff}=Q/3H(\rho_c+\rho_x)$.

\begin{figure*}[!h]
\includegraphics[width=6 cm]{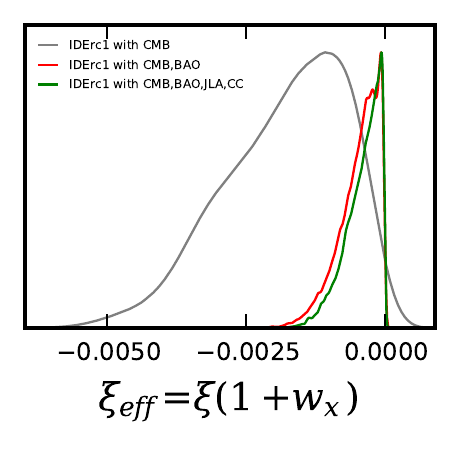} \quad \quad \quad 
\includegraphics[width=6 cm]{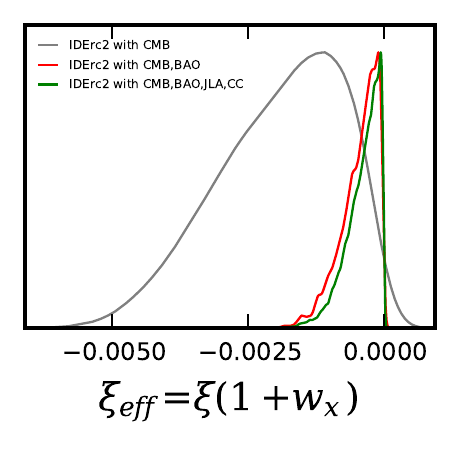}
\caption{1D marginalized posterior distribution of the parameter $\xi_{\rm eff}=\xi(1+w_x)$, which quantifies the effective DM-DE coupling strength and hence the direction of energy flow (from DE to DM if $\xi_{\rm eff}$ is negative, and from DM to DE if $\xi_{\rm eff}$ is positive), obtained by analyzing the CMB-only data (gray curves), the CMB+BAO dataset combination (red curves), and the CMB+BAO+JLA+CC dataset combination (green curves). The posteriors are plotted in arbitrary units and normalized to their maximums. The left panel is the posterior obtained assuming the \textit{IDErc1}, while the right panel is the posterior obtained assuming the \textit{IDErc2} model. The preference for negative values of $\xi_{\rm eff}$ indicates that energy flow occurs from the DM to the DE, as discussed in detail throughout the text.}
\label{1d_xieff}
\end{figure*}

Fig.~\ref{1d_xieff} shows the 1D marginalized posteriors of $\xi_{\rm eff}$ through the three data combinations carried out in this work, for the models  \textit{IDErc1} (left panel) and  \textit{IDErc2} (right panel). We determine $\xi_{\rm eff}$ as a derived parameter from the parametric space $w_x - \xi$. For model \textit{IDErc1} we find the following 95\%~C.L. bounds on $\xi_{\rm eff}$: $\xi_{\rm eff} = -0.0017^{+0.0016}_{-0.0025}$, $\xi_{\rm eff} = -0.0005^{+0.0005}_{-0.0010}$, and $\xi_{\rm eff} = -0.0005^{+0.0005}_{-0.0013}$ from the CMB, CMB+BAO, and CMB+BAO+JLA+CC dataset combinations respectively. Similarly, for model \textit{IDErc2}, we obtain the following 95\%~C.L. bounds on $\xi_{\rm eff}$ from the same dataset combinations in the same order as previously: $\xi_{\rm eff} = -0.0018^{+0.0018}_{-0.0025}$, $\xi_{\rm eff} = -0.0005^{+0.0005}_{-0.0010}$, and $\xi_{\rm eff} = -0.0005^{+0.0005}_{-0.0012}$. We see that for all the three dataset combinations, within 95\%~C.L. negative values of $\xi_{\rm eff}$ are inferred. This is clearly shown in Fig.~\ref{1d_xieff}.

As previously anticipated, we note that due to the preference for a phantom DE component ($w_x < -1$) from the observational constraints, the effective coupling $\xi_{\rm eff}$ is inferred to be negative (and consequently $Q < 0$) at a significance which reaches 95\%~C.L. in all data combinations analyzed. The physical interpretation of this result envisages a coupling scenario where DM decays into DE. Similar results, where $Q < 0$ is also preferred are obtained in e.g.~\cite{Salvatelli:2014zta, Eu}.

For comparison, we show in Tab.~\ref{tab:results_lcdm} 68\%~C.L. credible intervals for the CMB-only dataset, assuming the $\Lambda$CDM (second column from the left) and $w$CDM (where the 6 parameters of $\Lambda$CDM plus $w_x$ are varied, right-most column) models. We do this in order to understand the effect of introducing the \textit{IDErc1} and \textit{IDErc2} coupling on the inferred values of the other cosmological parameters. We find an important $>2\sigma$ shift of the cold dark matter physical density towards higher values, with an error bar twice as large, when the IDE models are considered (recall that these shifts are discussed for the CMB-only dataset).

An interesting shift, however, is that experienced by the Hubble parameter $H_0$. As already discussed above, we expect $\xi$ and $H_0$ to be negatively correlated due to their mutual effect on the angle at which the sound horizon at last scattering is projected, which affects the position of the first peak. We expect that as $\xi$ is increased, $H_0$ should decrease. This expectation is borne out by our analysis, since we see that smaller mean values with respect to the $\Lambda$CDM case (for which $H_0=67.27\pm0.66\ \rm{km \ s^{-1} \ Mpc^{-1} }$) are preferred (for the $w$CDM case we only have a lower limit on $H_0$). On the other hand for the \textit{IDErc1} and \textit{IDErc2} models we have $H_0 = 66.2^{+3.2}_{-2.9}\ \rm{km \ s^{-1} \ Mpc^{-1} }$ and $H_0 = 65.8^{+3.4}_{-3.2}\ \rm{km \ s^{-1} \ Mpc^{-1} }$ respectively. Although the mean value of the posterior in the \textit{IDErc1} and \textit{IDErc2} case is shifted to smaller values, the increase in the error bars implies that the tension with the locally determined value of $H_0=73.24\pm1.74\ \rm{km \ s^{-1} \ Mpc^{-1} }$ obtained by Riess \textit{et al.}~\cite{R16} is brought down to the level of $\sim 2\sigma$ (see also). In other words, the introduction of the DM-DE interaction has provided a channel for reducing the $>3\sigma$ tension between CMB and local determinations of $H_0$. We stress, however, that the reduction of the tension is not due to a genuine shift in the mean value of the posterior, but rather to an increase in the error bar by a factor of $\approx 5$, for the increasing of the volume of the parameter space.

It is furthermore interesting to examine how $\sigma_8$ shifts, in order to investigate if the IDE could alleviate the tension at more than $2\sigma$ present between the CMB and the weak lensing measurements from the CFHTLenS survey \cite{Heymans:2012gg, Erben:2012zw, Joudaki:2016mvz} and KiDS-450 \cite{Hildebrandt:2016iqg}. We find that this parameter moves towards lower values with respect to the $\Lambda$CDM case (unlike in the $w$CDM case where it increases), while its error bar increases by a factor of $2$. However, if we consider $S_8 \equiv\sigma_8 \sqrt{\Omega_m/0.3}$, i.e. the quantity measured by the weak lensing experiments, we find $S_8 =0.879\, \pm 0.030$ ($S_8 =0.881\, \pm 0.030$) for \textit{IDErc1} (\textit{IDErc2}) from the CMB only, that means a $2.7\sigma$ ($2.8\sigma$) tension with the KiDS-450 measurements for which $S_8=0.745\pm0.039$~\cite{Hildebrandt:2016iqg}. Even if for CMB+BAO we find a smaller value $S_8 =0.847\, \pm 0.016$ ($S_8 =0.847\, \pm 0.017$) for \textit{IDErc1} (\textit{IDErc2}), the tension is still at $2.4\sigma$ for both the cases, and also for CMB+BAO+JLA+CC.

Interestingly, as already mentioned before, a phantom nature for the DE EoS is inferred even in the presence of the DM-DE coupling, as already determined for the CMB-only case in the $w$CDM model, as well as in extended models  considered in the literature~\cite{DiValentino:2017iww,DiValentino:2015ola,DiValentino:2016hlg,DiValentino:2017zyq,Mortsell:2018mfj}.

Finally, let us discuss the results of the Bayesian model comparison analysis we have performed. We compute $\ln B_{ij}$ (where $i=$\textit{IDErc1} or \textit{IDErc2} and $j=\Lambda$CDM) in order to quantify the strength of the evidence for $\Lambda$CDM. In particular, given the definition of $\ln B_{ij}$, a negative value of this quantity corresponds to evidence for $\Lambda$CDM, and a positive value to evidence for the IDE model. We compute $\ln B_{ij}$ for the \textit{IDErc1} and \textit{IDErc2} models (as well as for $\Lambda$CDM) for the CMB, CMB+BAO and CMB+BAO+JLA+CC dataset combinations using the \texttt{MCEvidence} code. 

We find that, for the CMB only $\ln B_{ij} = -9.3$ for model \textit{IDErc1} and $\ln B_{ij} = -10.4$ for model \textit{IDErc2}. Using the revised Jeffreys scale of Tab.~\ref{tab:jeffreys}, in both cases this corresponds to a very strong preference for $\Lambda$CDM over the two IDE models. When adding BAO dataset we have smaller values, $\ln B_{ij} = -4.6$ for model \textit{IDErc1} and $\ln B_{ij} = -4.0$ for model \textit{IDErc2}, that indicate a strong preference for $\Lambda$CDM over the two IDE models. When considering the CMB+BAO+JLA+CC dataset, we find that the evidence for $\Lambda$CDM decreases further, since $\ln B_{ij} = -2.6$ for the \textit{IDErc1} model and $\ln B_{ij} = -2.4$ for the \textit{IDErc2} model. In both the cases, this corresponds to a positive preference for $\Lambda$CDM over the two IDE models. These results are summarized in Tab.~\ref{tab:bayesian}. In conclusion, even if the concordance $\Lambda$CDM model is always favored against the \textit{IDErc1} and \textit{IDErc2} models, we can see that the addition of further dataset decreases the $\Lambda$CDM evidence, with the strength of the evidence varying from very strong to strong to positive, depending on the dataset adopted (CMB, CMB+BAO+JLA+CC or CMB+BAO).

\begingroup                                                                                                                     
\begin{center}                                                                                                                  
\begin{table}[!h]                                                                                                                
\begin{tabular}{cccc}                                                                                                            
\hline\hline                                                                                                                    
Dataset & Model & $\ln B_{ij}$ & Strength of evidence for model $\Lambda$CDM \\ \hline
CMB & \textit{IDErc1} & $-9.3$ & Very Strong \\
CMB & \textit{IDErc2} & $-10.4$ & Very Strong \\
CMB+BAO & \textit{IDErc1} & $-4.6$ & Strong \\
CMB+BAO & \textit{IDErc2} & $-4.0$ & Strong \\
CMB+BAO+JLA+CC & \textit{IDErc1} & $-2.6$ & Positive \\
CMB+BAO+JLA+CC & \textit{IDErc2} & $-2.4$ & Positive \\
\hline\hline                                                                                                                    
\end{tabular}                                                                                                                   
\caption{Values of $\ln B_{ij}$, the logarithm of the Bayes factor for the two IDE models with respect to $\Lambda$CDM, obtained in our analysis for different dataset combinations, and corresponding strength of the evidence for $\Lambda$CDM qualified according to the modified Jeffreys scale given in Tab.~\ref{tab:jeffreys}. The negative values of $\ln B_{ij}$ indicate that $\Lambda$CDM is preferred over the two IDE models from the Bayesian evidence point of view.}\label{tab:bayesian}                                                                                                   
\end{table}                                                                                                                     
\end{center}                                                                                                                    
\endgroup 

\section{Summary and discussions}
\label{sec-discuss}

Theories where dark matter and dark energy interact play an important role in cosmology. Being initially motivated in order to solve the cosmological constant problem and later the coincidence problem, interacting dark energy theories have recently experienced a renewed surge of interest due to their ability of addressing the well-known tensions between high- and low-redshift estimates of $H_0$ and $\sigma_8$.

An important problem in interacting dark energy models is that these usually lead to early-time instabilities due to the fact that dark energy pressure perturbations are proportional to $(1+w_x)^{-1}$, where $w_x$ is the equation of state of dark energy. Therefore, for $w_x \rightarrow -1$, pressure perturbations blow up, leading to the blowing up of curvature perturbations on super-horizon scales. For this reason, observational tests of such models were typically forced to separate the $w_x$ parameter space into two discontinuous regions, namely the non-phantom ($w_x> -1$) and phantom regions ($w_x< -1$). Obviously, this separation of the parameter space leads to information loss and might obscure the possible observational viability of the model.

The problem outlined above can be solved by a simple transformation of the coupling parameter $\xi \rightarrow \xi_{\rm eff} = \xi (1+w_x)$~\cite{Yang:2017zjs,Yang:2017ccc}: this cancels the factor of $(1+w_x)^{-1}$ in the pressure perturbations, removing the potential instability and thus allowing one to study the entire $w_x$ parameter space without restrictions. We remark that the effective coupling parameter for these novel types of interactions, i.e. $\xi_{\rm eff}$, depends on the DE equation of state parameter $-$ as far as we aware, this feature has yet to be studied in the literature, if we exclude the recent works~\cite{Yang:2017zjs,Yang:2017ccc}. Therefore, undoubtedly, this kind of studies demand further investigations. In this work, along the lines of~\cite{Yang:2017zjs,Yang:2017ccc}, we have proposed two new interacting dark energy models (\textit{IDErc1} and \textit{IDErc2}). We have shown that the models are stable for $\xi >0$, independently of the values of $w_x$. We have then constrained these models in light of recent cosmological observations.

The observational constraints on the cosmological parameters of the two interacting dark energy models obtained from our analyses are summarized in Tab.~\ref{tab:results_1} and Tab.~\ref{tab:results_2}, as well as in Fig.~\ref{1D_1}, Fig.~\ref{1D_2}, Fig.~\ref{2D_BAO_1}, and Fig.~\ref{2D_BAO_2}. In particular, we found new stringent upper bounds on the strength of the DM-DE coupling strength. When using the full dataset combination (CMB+BAO+JLA+CC) we obtain $\xi < 0.016$ and $\xi < 0.015$ at 95\%~C.L. for the \textit{IDErc1} and \textit{IDErc2} models respectively (similar bounds are obtained from the CMB and CMB+BAO dataset combinations). On the other hand, we find that a phantom character of the dark energy equation of state is preferred, with significance surpassing 95\%~C.L. for the full dataset combination. As the effective coupling parameter $\xi_{\rm eff} = \xi(1+w_x)$ depends on $w_x$, although $\xi> 0$ is required for the stability of perturbations, the phantom character of $w_x$ implies that $1+w_x$ is negative and hence the coupling functions for both models become negative (i.e. $Q <0$, since $\rho_c>0$ and $(\rho_c +\rho_x) >0$ in the interaction functions, given by Eq.~(\ref{model1}) and Eq.~(\ref{model2}). The physical picture then corresponds to an energy flow from DE to DM or, depending on the fundamental physics model underlying the energy flow, a decay of DE field into DM.

An interesting observational concern is the effect of the DM-DE coupling on the inferred value of the Hubble parameter. We find that for both interacting models, when considering only CMB data, introducing the DM-DE coupling lowers the inferred value of the Hubble parameter (for the \textit{IDErc1} model $H_0=66.2^{+3.2}_{-2.9}\, \rm{km \ s^{-1} \ Mpc^{-1} }$, while for the \textit{IDErc2} model $H_0=65.8^{+3.4}_{-3.2}\, \rm{km \ s^{-1} \ Mpc^{-1} }$). This has to be compared with the value inferred assuming $\Lambda$CDM (for which $H_0=67.27\pm0.66\ \rm{km \ s^{-1} \ Mpc^{-1} }$)~\cite{Ade:2015xua}. Although the central value we obtain is lower, we notice that the error bars increase significantly. As a result of this, despite the decrease in the mean value, the value of $H_0$ we infer within the two interacting dark energy models is actually in better agreement with the local measurements of Riess \textit{et al.} ($H_0 = 73.24 \pm 1.74 \rm{km \ s^{-1} \ Mpc^{-1}}$)~\cite{R16} than the value inferred from $\Lambda$CDM by Planck \cite{Ade:2015xua}. Thus, in the interacting dark energy models we considered, the $H_0$ tension is alleviated. This is in agreement with the conclusions of similar recent works~\cite{Kumar:2016zpg, DiValentino:2017iww}. However, the $\sigma_8$ tension is not alleviated.


Finally, we performed a model comparison analysis where we computed the Bayes factors for the two interacting dark energy models with respect to the concordance $\Lambda$CDM model. The results are summarized in Tab.~\ref{tab:bayesian}. We find that $\Lambda$CDM is always preferred over the two interacting dark energy models, from the Bayesian evidence standpoint. The degree of preference varies from positive to strong, depending on the datasets adopted. In particular, we observe that the strength of the preference for $\Lambda$CDM decreases when moving from the CMB-only dataset, to the CMB+BAO, and then CMB+BAO+JLA+CC dataset combination.

In conclusion, we have constructed novel stable interacting dark energy models, wherein the strength of the DM-DE coupling explicitly depends on the dark energy equation of state, and the stability is guaranteed for any value of the latter. We have then constrained these interacting dark energy models against state-of-the art CMB and large-scale structure measurements, and found that the data indicate a scenario where energy flows from the DM to the DE field, due to the preference for phantom DE and the specific dependence of the coupling function on the DE equation of state. We have shown that the two models can alleviate the $H_0$ tension but not the $\sigma_8$ one. Finally, we have checked that the two models are not preferred over $\Lambda$CDM from the Bayesian evidence point of view.

Further work on these novel models is certainly warranted. There are in fact several interesting areas worth investigating, for instance the effect of including massive neutrinos into the physical picture under examination. The sum of the neutrino masses $M_{\nu}$ is tightly constrained when assuming a $\Lambda$CDM background~\cite{Cuesta:2015iho,Huang:2015wrx,DiValentino:2015sam,Giusarma:2016phn,Vagnozzi:2017ovm,Giusarma:2018jei,Nunes_Bonilla,Couchot:2017pvz,Doux:2017tsv,Wang:2017htc,Chen:2017ayg,Upadhye:2017hdl,Zennaro:2017qnp,Moresco:2016nqq,Salvati:2017rsn}, and thanks to prior volume effects these strong bounds are starting to provide interesting information on the mass ordering as well~\cite{Hannestad:2016fog,Xu:2016ddc,Gerbino:2016ehw,Gerbino:2016sgw,Simpson:2017qvj,Schwetz:2017fey,Hannestad:2017ypp,Long:2017dru,Gariazzo:2018pei,Capozzi:2017ipn,Heavens:2018adv,Handley:2018gel}, while the future appears bright with a detection realistically upcoming~\cite{Archidiacono:2016lnv,Allison:2015qca,DiValentino:2016foa,Sprenger:2018tdb,
Vagnozzi:2018pwo,Brinckmann:2018owf,Ade:2018sbj}. However, the picture is expected to change significantly when a different dark energy background is assumed~\cite{Kumar:2016zpg, Q3,Guo,Zhang:2015rha,Yang:2017amu,Wang:2016tsz,Vagnozzi:2018jhn,Kumar_Nunes_Yadav,Zhang:2015uhk,Zhang:2017rbg}. We hope to report the results in a companion paper after the upcoming final release of \textit{Planck} results.

\section*{ACKNOWLEDGMENTS}
The authors would like to thank the referee for very useful comments which improved the quality of our discussion. The authors would also like to thank Olga Mena and Alessandro Melchiorri for useful discussions. WY acknowledges  the support from the National Natural Science Foundation of China under Grants No. 11705079 and No. 11647153. EDV acknowledges support from the European Research Council in the form of a Consolidator Grant with number 681431. SV acknowledges support by the Vetenskapsr\r{a}det (Swedish Research Council) through contract No. 638-2013-8993 and the Oskar Klein Centre for Cosmoparticle Physics. DFM acknowledges the support from the Research Council of Norway, and this paper is based upon work from COST action CA15117 (CANTATA), supported by COST (European Cooperation in Science and Technology).

\end{document}